\documentclass{aa}

\usepackage{graphicx} 
\usepackage{txfonts}
\usepackage{natbib}
\bibpunct{(}{)}{;}{a}{}{,} 

\begin{document}

\title{Measuring the fading of S0 galaxies using globular clusters}
\author{A. Arag\'on-Salamanca \and A.G. Bedregal \and M.R. Merrifield}
\institute{School of Physics and Astronomy, University of Nottingham,
  NG7 2RD, UK}
\date{Received *** / Accepted ***}

\abstract
{}
{
We test the hypothesis that S0 galaxies are the
descendants of fading spirals whose star formation has been shut down,
by using the properties of their globular cluster systems. 
} 
{
We estimate the amount by which the globular cluster specific
frequency (number of globular clusters per unit $V$-band luminosity)
is enhanced in S0s relative to spirals.  If the transformation
hypothesis is correct, and no clusters are created or destroyed in the
process, then this difference provides a measure of the
degree to which the S0's $V$-band luminosity has faded relative to that
of its spiral progenitor, which we can compare with the independent
values estimated from stellar population synthesis and the S0
Tully--Fisher relation.  We also explore whether the degree to
which the globular cluster specific frequency is enhanced in S0s
correlates with the colour of the stellar population, as also
predicted by this hypothesis in which galaxies become redder as they
fade.
}
{
We find that, on average, the globular cluster specific frequency is a
factor $\sim 3$ larger for S0s than for spirals, which can be
interpreted as meaning that passively-evolving S0s have faded on
average by about a factor of three from their spiral progenitors.
This value fits remarkably well with the predictions of stellar
population synthesis calculations, and the offset between the S0 and
spiral Tully--Fisher relations, where the S0 $V$-band relation lies
$\sim 1.2$ magnitudes, or a factor of three, below the spiral
relation.  We also find that the global colours of S0 galaxies are
strongly correlated with their globular cluster specific frequencies:
the redder the stellar population of an S0, the larger its specific
frequency, as we might expect if we are catching different S0s at
different stages of passively fading and reddening.  Comparison to the
predictions of stellar population synthesis models show that this
explanation works quantitatively as well as qualitatively.
}
{
These tests strongly support the hypothesis that S0 galaxies were
once normal spirals, whose star formation was cut off, presumably due
to a change of environment.  We are now in a position to start to make
quantitative measurements of when this life-changing event occurred in
different galaxies.
}  

\maketitle

\section{Introduction}
S0, or lenticular, galaxies live at the crossroads between elliptical
and spiral galaxies in the traditional Hubble tuning-fork diagram,
suggesting that they should play a key role in understanding the
morphologies of galaxies.  In fact, in some cluster environments, S0s
are the single most common type of luminous galaxy
\citep{Dressler:1980}, so clearly understanding how they form and
evolve is essential if we wish to have a complete picture of how
galaxy morphology is related to galaxy formation and the environment.
\citet{Dressler:1980} also showed that as the fraction of S0s
increases in the densest cluster environments, so the fraction of
spirals decreases, naturally suggesting that S0s are simply ``dead''
spiral galaxies that have had their star formation shut off by their
surroundings, and are now quietly fading away as their stellar
populations age.  Indeed, a number of mechanisms have been suggested
that might cause such a transformation: ram-pressure stripping of disk
gas could remove the raw material of the next stellar generation
\citep{Gunn_Gott:1972}, while the somewhat gentler removal of a
larger-scale reservoir of halo gas could lead to a slower
``strangulation'' of star formation \citep{Larson_etal:1980}.

Although this scenario is quite plausible, we need some more direct
evidence that it has actually occurred.  Circumstantial evidence for
such a transformation over cosmic timescales comes from observations
which indicate that the proportion of S0 galaxies is substantially
smaller in distant clusters than in nearby ones, while spirals show
the opposite trend \citep[see, for example,][]{Dressler_etal:1997}, but any
inferences drawn from such studies always face the criticism that one
may not be equating comparable systems at different redshifts.
Further evidence that S0s are fading spirals comes from studies of the
Tully--Fisher relation between luminosity and rotation speed for S0
galaxies, which is found to be rather broad and offset from the
relation for spirals in the sense that the S0s are systematically
fainter than the spirals.  A recent analysis by
\citet{Bedregal_etal:2006} found that the S0 relation in the $B$-band
is offset by $\sim 1.4$ magnitudes from the \citet{Sakai_etal:2000}
relation for spirals, with a scatter of $\sim 1$ magnitude.  Such an
offset can most straightforwardly be interpreted as arising from the
fading of the S0s relative to their spiral progenitors, with the
spread in the relation arising from the different epochs at which this
fading commenced.  Indeed, \citet{Bedregal_etal:2006} were able to
uncover some evidence that the magnitude of each S0's offset from the
spiral Tully--Fisher relation depends on the time since star
formation was cut off, as measured by the age of its stellar
population, just as this picture would predict.  However, the evidence
is still uncomfortably circumstantial, as we have had to assume that
the ancient progenitors of these S0s respect the same Tully--Fisher
relation that we see in the nearby spirals, and that the difference
between the relations for S0s and spirals arises from luminosity
evolution: the offset could also arise because these systems have
fundamentally different mass properties, shifting the relation in
rotation speed.  Even with these assumptions, we can only tie spirals
and S0s together on a statistical basis, as we do not have a direct
connection between any S0 and the properties of its individual
progenitor.

Ideally, we would like to find some historical record preserved in the
current properties of an S0 that tells us about its luminosity during
its early life as a spiral, and that will not have been defaced by the
transformation process or any subsequent evolution of the galaxy.
Fortunately, just such a record exists in the globular cluster (GC)
population of the S0.  As we will see below, GCs have a reasonably well
defined specific frequency (number per unit galaxy $V$-band
luminosity) for spiral galaxies, so the number of GCs provides a
useful proxy for a spiral galaxy's luminosity.  Further, the
transformation from spiral to S0 is unlikely to alter the number of
GCs significantly: the hydrodynamic processes that strip out the gas
from the spiral will not have any impact on these collisionless
stellar clusters, and the conversion is likely to be sufficiently
benign that we are unlikely to lose clusters or produce any new ones
as we find in more dramatic phenomena like mergers
\citep{Ashman_and_Zepf:1998}.  In addition, the GCs are sufficiently
old that their own passive fading will be very slow, so the observed
number counts should not decrease significantly during the
transformation.  Thus, the number of GCs in an S0 offers a reasonably
robust indicator of its spiral progenitor's luminosity, so the
specific frequency of GCs in such a system provides a direct measure
of the ratio of its progenitor's luminosity to its current luminosity,
which we can compare to the less direct indicators of fading such as
those provided by the Tully--Fisher relation.

In this paper, we carry out such an analysis, comparing the
specific frequencies of GCs in S0s to those in spirals in
Section~\ref{sec:GCSFcomp} to see how much these systems must have
faded, and going on to look for evidence that different amounts of
fading might be related to different epochs of transformation in
Section~\ref{sec:fadeage}.  Conclusions are presented in
Section~\ref{sec:conc}.

\section{The specific frequency of globular clusters in S0s and
  spirals}\label{sec:GCSFcomp}

The specific frequency of GCs in a galaxy, $S_N$, is
usually defined relative to the galaxy's $V$-band luminosity,
normalized to $M_V=-15$:
\begin{equation} \label{eq:SN}
S_N=N_{\rm t}10^{-0.4(M_V^{\rm T}+15)} 
\end{equation} 
where $N_{\rm t}$ is the total number of GCs and $M_V^{\rm T}$ is the
total absolute magnitude of the underlying galaxy
\citep{Harris_and_vandenBergh:1981}.  Since $S_N\propto N_{\rm
t}/L_V$, as discussed above, we would expect $S_N$ to grow as spiral
galaxies fade into S0s, if, indeed, they do.  Thus, by comparing the
specific frequency of GCs in spirals and S0s, it should be possible to
check whether $S_N$ is larger in S0s than in spirals as this scenario
requires, and to estimate by how much on average spirals fade during
this transformation, for comparison with less direct indicators like
the Tully--Fisher relation.

Unfortunately, data sets of GC specific frequencies in galaxies with
different morphologies do not exist for any reasonably-complete sample
of galaxies.  However, two entirely independent compilations have been
constructed, so we can make some assessment of the likely systematic
uncertainties that may arise from any selection biases by comparing
the results from these two data sets.

\begin{table}
\caption{Spiral and S0 absolute magnitudes and global globular cluster
 specific frequencies from \citet{Ashman_and_Zepf:1998}} 
\label{table:1} 
\centering 
\begin{tabular}{l c c l c c} 
\hline\hline 
\multicolumn{3}{c}{Spiral Galaxies} & \multicolumn{3}{c}{S0 Galaxies}\\ 
Name & $M_V^T$ & $S_N$ & Name & $M_V^T$ & $S_N$ \\
\hline 
MW         & $-21.3$ & $0.5\pm0.1$ & NGC 524    & $-22.5$ & $3.3\pm1.0$ \\
LMC        & $-18.4$ & $0.6\pm0.2$ & NGC 1387   & $-20.4$ & $2.7\pm0.8$ \\
M31        & $-21.8$ & $0.9\pm0.2$ & NGC 1553   & $-21.3$ & $1.4\pm0.3$ \\
M33        & $-19.4$ & $0.5\pm0.2$ & NGC 3115   & $-21.3$ & $1.6\pm0.4$ \\
NGC 253    & $-20.2$ & $0.2\pm0.1$ & NGC 3384   & $-20.4$ & $0.9\pm0.5$ \\
NGC 3031   & $-21.1$ & $0.7\pm0.1$ & NGC 3607   & $-20.9$ & $3.5\pm2.6$ \\
NGC 2683   & $-20.8$ & $1.7\pm0.5$ & NGC 4340   & $-19.9$ & $8.5\pm3.1$ \\
NGC 4565   & $-21.5$ & $0.5\pm0.1$ & NGC 4526   & $-21.4$ & $7.4\pm2.2$ \\
NGC 4216   & $-21.8$ & $1.2\pm0.6$ \\
NGC 4569   & $-21.7$ & $1.9\pm0.6$ \\
NGC 4594   & $-22.2$ & $2.0\pm1.0$ \\
NGC 5170   & $-21.6$ & $0.9\pm0.3$ \\
NGC 7814   & $-20.4$ & $3.5\pm1.1$ \\
\hline
Average    &         & $1.2\pm0.3$ &            &         & $3.7\pm1.1$ \\
\hline 
\end{tabular}
\end{table}

\begin{figure*}
\makebox[\textwidth]{
    \includegraphics[width=0.80\textwidth]
      {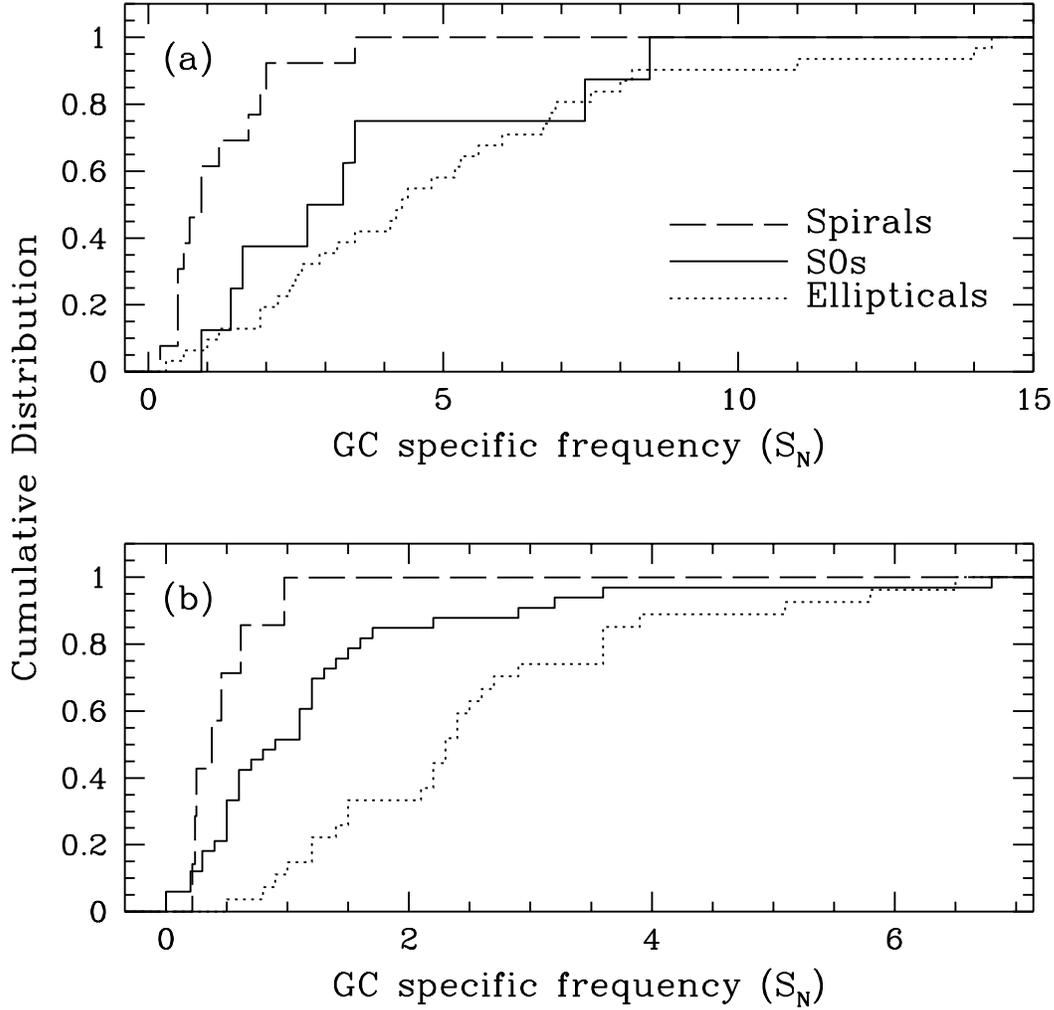}
}
\caption 
{
(a) Cumulative distribution of global GC specific frequencies for
spiral, S0 and elliptical galaxies from \citet{Ashman_and_Zepf:1998}. (b)
Cumulative distribution of local GC specific frequencies for
spiral, S0 and elliptical galaxies from Kundu \& Whitmore (2001a,b)
and \citet{Goudfrooij_et_al:2003}.
}
\label{fig1}
\end{figure*}

First, we have analyzed the values published by
\citet{Ashman_and_Zepf:1998}, excluding dwarf galaxies ($M_V>-17$) and
cDs.  The relevant data for spirals and S0s are summarized in
Table~\ref{table:1}.  In Figure~\ref{fig1}(a), we plot the cumulative
distributions of GC specific frequencies determined for spirals and
S0s. For completeness, we also show the distribution for elliptical
galaxies, but these have no direct bearing on the current analysis.
It is immediately apparent that S0s have, on average, significantly
larger specific frequencies than spirals. A Kolmogorov--Smirnov test
rejects the hypothesis that spirals and S0s have the same $S_N$
distribution at the 90\% confidence level, and we find that for this
sample
\begin{equation}\label{eq:comp1}
\frac{\langle S_N^{\rm S0}\rangle}{\langle S_N^{\rm Sp}\rangle} = 3.1\pm1.1
\end{equation} 
where $\langle S_N^{\rm Sp}\rangle$ and $\langle S_N^{\rm S0}\rangle$
are the unweighted average GC specific frequencies for spirals and S0s
respectively.  The uncertainties on the specific frequencies were
calculated from the standard error in the mean, and these
uncertainties were propagated to provide the quoted error in
equation~\ref{eq:comp1}.

\begin{table*}
\caption{Absolute magnitudes and globular cluster
 specific frequencies from \citet{Kundu_and_Whitmore:2001b} for S0s, and
\citet{Goudfrooij_et_al:2003} for spiral galaxies, with global colours
for S0s where available.} 
\label{table:2} 
\centering 
\begin{tabular}{l c c c c c c c} 
\hline\hline 
Name & $M_V^T$ & $S_N^{\rm global}$ & $S_N^{\rm local}$ & $(U-B)$ & $(B-V)$ & $(V-R)$ & $(V-I)$  \\
\hline 
\phantom{\tiny a}\\
\multicolumn{8}{c}{Spiral Galaxies}\\
NGC 3628     &   $-21.0$ &   $1.9\pm0.2$   & $1.0\pm0.1$  & -- & -- & -- & --  \\
NGC 4013     &   $-20.8$ &   $1.1\pm0.3$   & $0.6\pm0.2$  & -- & -- & -- & --  \\
NGC 4517     &   $-21.6$ &   $0.6\pm0.2$   & $0.2\pm0.1$  & -- & -- & -- & --  \\
NGC 4565     &   $-21.5$ &   $0.6\pm0.2$   & $0.2\pm0.1$  & -- & -- & -- & --  \\
NGC 4594     &   $-22.2$ &   $1.7\pm0.6$   & $0.4\pm0.2$  & -- & -- & -- & --  \\
IC 5176      &   $-21.1$ &   $0.5\pm0.1$   & $0.2\pm0.1$  & -- & -- & -- & --  \\
NGC 7814     &   $-20.5$ &   $0.7\pm0.2$   & $0.5\pm0.2$  & -- & -- & -- & --  \\
\hline
Average      &           &                 & $0.4\pm0.1$  &    &    &    &     \\
\hline
\phantom{\tiny a}\\
\multicolumn{8}{c}{S0 Galaxies}\\
NGC 524      &   $-22.4$ & --   &   $1.1\pm0.4$  & -- & -- & -- & -- \\
NGC 2768     &   $-22.0$ & --   &   $1.2\pm0.4$  & -- & -- & -- & -- \\
NGC 6861     &   $-21.8$ & --   &   $3.6\pm1.6$  &    $0.64\pm0.03$ & $1.04\pm0.03$ & $0.61\pm0.03$ & $1.31\pm0.03$  \\
NGC 6703     &   $-21.8$ & --   &   $1.2\pm0.8$  & -- & -- & -- & -- \\
NGC 1553     &   $-21.5$ & --   &   $0.5\pm0.1$  &    $0.49\pm0.03$ & $0.95\pm0.03$ & $0.57\pm0.03$ & $1.19\pm0.03$  \\
NGC 474      &   $-21.4$ & --   &   $0.7\pm0.5$  & -- & -- & -- & -- \\
NGC 3115     &   $-21.3$ & --   &   $1.3\pm0.1$  &    $0.59\pm0.03$ & $0.99\pm0.03$ & $0.61\pm0.03$ & $1.27\pm0.03$  \\
NGC 1332     &   $-21.2$ & --   &   $2.2\pm0.7$  &    $0.60\pm0.03$ & $1.03\pm0.03$ & $0.60\pm0.03$ & $1.30\pm0.03$  \\
NGC 3414     &   $-21.0$ & --   &   $1.6\pm0.6$  &    $0.55\pm0.01$ & $0.87\pm0.01$ & $0.70\pm0.01$ &          --    \\
NGC 4459     &   $-20.9$ & --   &   $0.9\pm0.3$  &    $0.47\pm0.01$ & $0.96\pm0.01$ & $0.55\pm0.01$ & $1.22\pm0.02$  \\
NGC 1201     &   $-20.8$ & --   &   $1.1\pm0.5$  &    $0.54\pm0.03$ & $0.97\pm0.03$ & $0.57\pm0.03$ & $1.22\pm0.03$  \\
NGC 1400     &   $-20.6$ & --   &   $2.9\pm1.1$  &    $0.61\pm0.03$ & $1.03\pm0.03$ & $0.65\pm0.03$ & $1.37\pm0.03$  \\
NGC 3607     &   $-20.4$ & --   &   $1.3\pm0.4$  & -- & -- & -- & -- \\
NGC 4203     &   $-20.2$ & --   &   $1.5\pm0.5$  & -- & -- & -- & -- \\
NGC 2902     &   $-20.2$ & --   &   $0.3\pm0.4$  & -- & -- & -- & -- \\
NGC 3489     &   $-19.6$ & --   &   $1.4\pm0.7$  & -- & -- & -- & -- \\
NGC 4379     &   $-19.6$ & --   &   $0.6\pm0.4$  &    $0.39\pm0.01$ & $0.85\pm0.01$ & $0.61\pm0.01$ & $1.24\pm0.02$  \\
NGC 1389     &   $-19.5$ & --   &   $0.5\pm0.4$  &    $0.43\pm0.01$ & $0.92\pm0.01$ & $0.56\pm0.01$ & $1.16\pm0.02$  \\
NGC 3056     &   $-18.9$ & --   &   $0.6\pm0.7$  & -- & -- & -- & -- \\
NGC 3156     &   $-18.9$ & --   &   $0.4\pm0.4$  & -- & -- & -- & -- \\
IC 3131      &   $-18.9$ & --   &   $0.2\pm0.4$  & -- & -- & -- & -- \\
NGC 1375     &   $-18.8$ & --   &   $0.5\pm0.7$  &    $0.31\pm0.03$ & $0.81\pm0.03$ & $0.51\pm0.03$ & $1.06\pm0.03$  \\
VCC 165      &   $-18.5$ & --   &   $0.0\pm2.1$  & -- & -- & -- & -- \\
NGC 3599     &   $-18.4$ & --   &   $1.2\pm0.9$  & -- & -- & -- & -- \\
NGC 2328     &   $-18.4$ & --   &   $0.0\pm1.4$  & -- & -- & -- & -- \\
NGC 4431     &   $-18.4$ & --   &   $0.5\pm0.7$  & -- & -- & -- & -- \\
IC 1919      &   $-18.2$ & --   &   $1.1\pm1.7$  & -- & -- & -- & -- \\
NGC 1581     &   $-18.2$ & --   &   $0.2\pm0.9$  &    $0.20\pm0.03$ & $0.73\pm0.03$ & $0.46\pm0.03$ & $1.06\pm0.03$  \\
ESO 358-G059 &   $-18.0$ & --   &   $0.6\pm1.3$  & -- & -- & -- & -- \\
NGC 3870     &   $-17.9$ & --   &   $0.8\pm1.3$  & -- & -- & -- & -- \\
NGC 3115-DW1 &   $-17.9$ & --   &   $6.8\pm2.4$  & -- & -- & -- & -- \\
ESO 118-G034 &   $-17.6$ & --   &   $0.3\pm2.0$  & -- & -- & -- & -- \\
IC 3540      &   $-17.3$ & --   &   $3.2\pm4.2$  & -- & -- & -- & -- \\
NGC 4150     &   $-16.6$ & --   &   $1.7\pm1.8$  & -- & -- & -- & -- \\
\hline
Average      &           &      &   $1.2\pm0.2$  &    &    &    &     \\
\hline 
\end{tabular}
\end{table*}
We have constructed the second $S_N$ data set using the results
published by \citet{Kundu_and_Whitmore:2001a} for elliptical galaxies,
\citet{Kundu_and_Whitmore:2001b} for S0s, and
\citet{Goudfrooij_et_al:2003} for spiral galaxies.  Again, dwarf
galaxies have been excluded.  The adopted data for spirals and S0s are
presented in Table~\ref{table:2}.  One complication here is that Kundu
and Whitmore published mainly ``local'' $S_N$ values, corresponding to
one WFPC2 field, while Goudfrooij et al.\ give global values
comparable to those described above.  However, Kundu \& Whitmore
(2001a,b, hereafter KW) give both local and global specific
frequencies for 14 galaxies, so it is possible to calibrate the
difference between these quantities, and how they might depend on the
galaxies' absolute magnitude.  A least-squares fit to these data
yields a calibration of
\begin{equation}
\frac{S_N^{\rm local}}{S_N^{\rm global}}=
5.71 + 0.25M_V^{\rm T}
\end{equation}
with a scatter of only 0.18.  Since we are primarily interested in
population average values of $S_N$, and we are looking at variations
between populations that amount to factors of several, this $\sim
20$\% uncertainty is completely negligible.  Ideally, we would use
this transformation to convert the local values from KW into global
values.  However, the calibration of this relation depends almost
entirely on galaxies with absolute magnitudes $M_V < -20$, and a
number of galaxies in the KW sample are fainter than this limit, so
such a transformation would have significant uncertainties.
Fortunately, all the spiral galaxies in the
\citet{Goudfrooij_et_al:2003} sample lie at brighter magnitudes, so we
can use the relation to transform their quoted global values into
local values, as shown in Table~\ref{table:2}; since we are only
interested in the ratio of specific frequencies, it does not matter
whether the local or global values are employed as long as we use the
same in both samples.  In figure~\ref{fig1}(b) we plot the cumulative
distributions of GC local specific frequencies for spirals, S0s and
ellipticals from this second compilation. Again, S0s have, on average,
significantly larger specific frequencies than spirals (once again, a
Kolmogorov--Smirnov test rejects the hypothesis that spirals and S0s
have the same $S_N$ distribution at the $90$\% confidence level,
giving a rejection from the combined data sets that is significant at
the 99\% level).  For this second dataset, we find that the ratio of
local specific frequencies is
\begin{equation}\label{eq:comp2}
\frac{\langle S_N^{\rm S0}\rangle}{\langle S_N^{\rm Sp}\rangle}
\simeq2.7\pm0.9.
\end{equation}
Given that equations \ref{eq:comp1} and \ref{eq:comp2} were derived
from completely different data sets derived using different
instruments using different analysis techniques applied by different
authors, the agreement between the two values suggests that this
result is reasonably robust, notwithstanding the incompleteness of the
samples.  It would appear that spirals have a specific frequency of GCs
that is a factor of $\sim 3$ lower than that in S0s, with the
fairly simple interpretation that the latter have held onto their GCs
but faded by a factor of $\sim 3$ in transforming from the former.  

Is this factor plausible?  Well, population synthesis models based on
\citet{ Bruzual_and_Charlot:2003} indicate that a galaxy with an
initially steady star formation rate will fade by a factor of $\sim3$
in the $V$~band in $\sim3\,$Gyr after complete truncation of its star
formation.  If the star formation decays more slowly (as through the
strangulation mechanism), such fading could take $\sim4$--$6\,$Gyr
[see \citet{Shioya_etal:2004} for some model examples].  These
timescales are comparable to the look-back time to $z\sim0.5$, which
is where \citet{Dressler_etal:1997} found signs of the transition from
spiral-dominated to S0-dominated clusters. However, we would not
expect all galaxies to undergo the transition at the same time, and
presumably some fraction of the significant scatter in $S_N$ for S0s
can be attributed to the different amount of fading that has taken
place for different systems that underwent this transformation at
different times, a point to which we return in
Section~\ref{sec:fadeage}. Nonetheless, it is heartening that the
difference in the mean value of $S_N$ seems entirely consistent with
the characteristic look-back time at which the transformation is
believed to have predominantly occurred.

The difference also matches up to the results from the analysis of the
Tully--Fisher relation.  As mentioned above, \citet{Bedregal_etal:2006}
found a $\sim 1.4$ magnitude offset between the $B$-band Tully--Fisher
relation for S0s and that for spirals.  Using the same
passively-evolving stellar population models as described above, we
find that such a fading in the $B$-band corresponds to a fading in the
$V$-band of $\sim 1.2$ magnitudes, or a factor of $\sim 3$ decrease in
$V$-band luminosity, just as found in the analysis of globular
clusters.  Although most reassuring, the exact agreement must be to an
extent coincidental: \citet{Bedregal_etal:2006} found significant
uncertainty in the offset in the Tully--Fisher relations depending on
the exact calibration adopted, so the factor of $\sim 3$ predicted by
the Tully--Fisher analysis could easily be a factor of two or four.
The lack of a complete sample of S0s upon which to carry out this test
also means that some residual uncertainty must remain in the absolute
value of the mean fading of the population, due to unknown systematic
effects in the way that the sample was selected.

Thus, although the mean degree of fading as derived from the globular
cluster specific frequency is entirely consistent with the values
predicted by population synthesis and those derived independently from
the Tully--Fisher relation, the case would be even stronger if there
were a more subtle test that goes beyond simple absolute shifts in the
mean of the population.

\section{Correlation between globular cluster specify frequency and 
galaxy colours}\label{sec:fadeage}

Happily, there is a differential test that we can apply to the GC
specific frequency data that overcomes these uncertainties in both
absolute magnitude and sample selection.  As an individual system
fades from spiral to S0, the specific frequency of GCs will go up as
we have discussed above, and the colours of the galaxy should redden
as the stellar population ages.  Thus, we should expect to see a
correlation between the measured values of $S_N$ and the S0s' colours,
with different values corresponding to the degree to which the S0 has
evolved away from its spiral progenitor.  Further, we can compare the
form of this relation to that predicted by the population synthesis
calculations to see whether the correlation matches the model
predictions in detail.

Unfortunately, the necessary data do not exist for all of the S0
galaxies for which $S_N$ values are available.  The evolution in
colour in such a fading population is relatively modest, so accurate
photometry is required, and we are interested in the galaxies' global
stellar populations, so need spatially-resolved photometry rather than
just central values for colours.  Such high-quality extended data are
not readily available for many galaxies, but using the photometry of
\citet{Poulain_and_Nieto:1994} and the highest-quality data from the
compilation of \citet{Prugniel_and_Heraudeau:1998}, we have been able
to bring together reliable multi-colour photometry for 12 of the
non-dwarf S0 galaxies from \citet{Kundu_and_Whitmore:2001b}, which we
summarize in Table~\ref{table:2}. These 12 galaxies appear to be a
fair sub-sample of the original sample, in that there is no obvious
bias in magnitude or the values of $S_N$.  To obtain a measure of the
global populations in these galaxies without unduly compromising the
photometric accuracy, we have calculated the integrated colours for
these systems within an aperture that corresponds to the $B$-band
half-light radius of each galaxy, as tabulated in \citet{RC2_mine}.

\begin{figure*}
\makebox[\textwidth]{
    \includegraphics[width=0.80\textwidth]
      {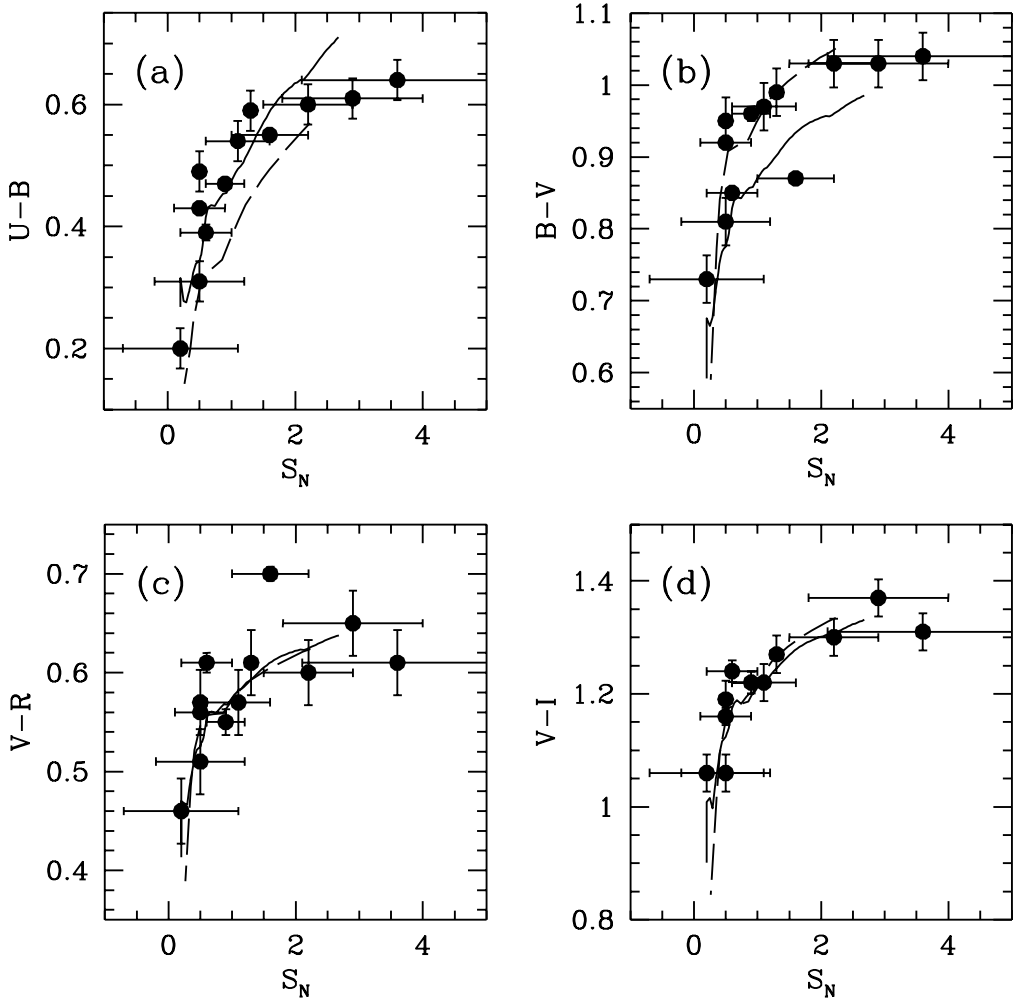}
}
\caption 
{ Global $U-B$, $B-V$, $V-R$ and $V-I$ colours versus globular cluster
specific frequency for S0 galaxies from
\citet{Kundu_and_Whitmore:2001b}.  The solid and and dashed lines
correspond to population synthesis model predictions of fading stellar
populations from \citet{Bruzual_and_Charlot:2003} and
\citet{Worthey:1994} respectively.  }
\label{fig2}
\end{figure*}

Figure~\ref{fig2} shows the derived global $U-B$, $B-V$, $V-R$ and
$V-I$ colours for these 12 S0 galaxies plotted against their GC
specific frequencies.  A clear correlation is observed in the sense
predicted by the fading spiral hypothesis: the larger the GC specific
frequency in an S0, the redder its stellar population.  

Further, the relation is of exactly the right amplitude.
Figure~\ref{fig2} shows the predictions of population synthesis codes
for a fading model stellar population with solar metallicity and a
standard initial mass function, which ceased forming stars more than
$\sim1$Gyr ago -- after this time, the colour evolution is not very
sensitive to the detailed history of the star formation that took
place before star formation ended, so the model predictions are fairly
robust.  To assess the uncertainty in the predictions, we computed
models using the codes of both \citet{Bruzual_and_Charlot:2003} and
\citet{Worthey:1994}, but, as is clear from Figure~\ref{fig2}, these
uncertainties are also fairly negligible compared to the error bars on
the data.  In these calculations, the variation in colour with age
comes directly from the models, and the changes in $S_N$ come from the
corresponding variations in the model values of the $V$-band
magnitude, using equation~\ref{eq:SN} and following the basic precept
of this paper that the number of GCs does not change with time.  The
$x$-axis zero-point of the models is fixed so that for the typical
$V-I$ colour of a star-forming spiral, $(V-I)\simeq1.1$
\citep{Fukugita_etal:1995}, the models predict $S_N\simeq0.4$, which
is the average value of $S_N^{\rm local}$ for the star-forming spirals
in the \citet{Goudfrooij_et_al:2003} sample (see Table~\ref{table:2}).
Once we have made this calibration to the local spirals, the
zero-points of the $x$-axes in all the other plots, the amplitude of
all the colours' variations with $S_N$, their slopes, and even their
curvatures are entirely fixed by the stellar evolution models, and, as
Figure~\ref{fig2} attests, the match of all these properties to the
observations is remarkably good.

\section{Conclusions}\label{sec:conc}

The classic problem in trying to address questions related to galaxy
evolution is that our snapshot view of the Universe means that we
never get to view the process directly.  Thus, although it seems quite
plausible that spiral galaxies fade to become S0s when their
interstellar media are stripped out and star formation ceases, there
is little direct evidence to this effect.  Here, we finesse this issue
by making use of the historical record imprinted in a galaxy via its
GC population, which should still be ``readable'' even after the
transformation process.  In particular, if the number of globular
clusters is taken as a measure of the progenitor spiral galaxy's
luminosity, and this number does not change as the galaxy passively
evolves, then we have been able to show that S0s must have faded by an
average factor of approximately three during this process.  This
result matches very neatly with the values predicted by stellar
population synthesis models and those found by comparing the
Tully--Fisher relations of spirals and S0s.

Further, using the amount by which the stellar population has
reddened as a measure of the time since the transformation into an S0
began, we have been able to show that individual galaxies are at
different stages along this evolutionary track: the colours of these
galaxies correlate with the specific frequency of GCs in exactly the
way that one would expect if their star formation shut down at
different times in the past.  Thus, we are at the point of being able
to follow the histories of individual S0 galaxies with the detail
necessary to say which underwent their life-changing transformations
first.

The next logical step in writing the life histories of these galaxies
is to obtain uniform samples of spectra of the quality necessary to
derive stellar population ages for these systems
\citep[e.g.][]{Kuntschner_Davies:1998}.  Not only would these data
resolve any remaining ambiguity by lifting the degeneracy between age
and metallicity effects in the interpretation of the broad-band
colours used here, but they would also allow us to put much more
accurate dates to the ``birthdays'' of S0 galaxies.

\begin{acknowledgements}
We would like to thank Steven Bamford, Jean Brodie, Bo Milvang-Jensen,
Osamu Nakamura and Frazer Pearce for very stimulating discussions.  We
would also like to thank the referee for suggestions that
substantially improved the paper.  MRM gratefully acknowledges the
support of a PPARC Senior Fellowship.
\end{acknowledgements}



\end{document}